\documentclass[%
 reprint, prx,
superscriptaddress,
 amsmath,amssymb,
 aps,
]{revtex4-2}
\usepackage{graphicx}
\usepackage{dcolumn}
\usepackage{bm}
\usepackage{textcomp} 
\usepackage{gensymb}
\usepackage{xcolor}
\usepackage{float}
\usepackage{physics}
\usepackage{siunitx}
\DeclareSIUnit{\Er}{E_{\text{rec}}}
\usepackage{hyperref}
\usepackage{tcolorbox}
\DeclareUnicodeCharacter{2212}{-}

\usepackage[normalem]{ulem}

\newcommand{\PR}[1][]{\frac{Vp^2}{U_0}}

\newcommand{\VTP}[1][]{V_{\text{TP}}}

\begin{document}

\preprint{APS/123-QED}

\title{Synchronization of Quasi-Particle Excitations in a Quantum Gas with Cavity-Mediated Interactions}
\author{Gabriele Natale}
\thanks{These authors contributed equally to this work.}
\affiliation{Institute for Quantum Electronics, Eidgen\"ossische Technische Hochschule Z\"urich, Otto-Stern-Weg 1, CH-8093 Zurich, Switzerland}

\author{Alexander Baumgärtner}
\thanks{These authors contributed equally to this work.}
\affiliation{Institute for Quantum Electronics, Eidgen\"ossische Technische Hochschule Z\"urich, Otto-Stern-Weg 1, CH-8093 Zurich, Switzerland}

\author{Justyna Stefaniak}
\affiliation{Institute for Quantum Electronics, Eidgen\"ossische Technische Hochschule Z\"urich, Otto-Stern-Weg 1, CH-8093 Zurich, Switzerland}

\author{David Baur}
\affiliation{Institute for Quantum Electronics, Eidgen\"ossische Technische Hochschule Z\"urich, Otto-Stern-Weg 1, CH-8093 Zurich, Switzerland}

\author{Simon Hertlein}
\affiliation{Institute for Quantum Electronics, Eidgen\"ossische Technische Hochschule Z\"urich, Otto-Stern-Weg 1, CH-8093 Zurich, Switzerland}

\author{Dalila Rivero}
\affiliation{Institute for Quantum Electronics, Eidgen\"ossische Technische Hochschule Z\"urich, Otto-Stern-Weg 1, CH-8093 Zurich, Switzerland}

\author{Tilman Esslinger}
\affiliation{Institute for Quantum Electronics, Eidgen\"ossische Technische Hochschule Z\"urich, Otto-Stern-Weg 1, CH-8093 Zurich, Switzerland}

\author{Tobias Donner}
\email{donner@phys.ethz.ch}
\affiliation{Institute for Quantum Electronics, Eidgen\"ossische Technische Hochschule Z\"urich, Otto-Stern-Weg 1, CH-8093 Zurich, Switzerland}

\date{\today}

\begin{abstract}
Driven-dissipative quantum systems can undergo transitions from stationary to dynamical phases, reflecting the emergence of collective non-equilibrium behavior. We study such a transition in a Bose-Einstein condensate coupled to an optical cavity and develop a cavity-assisted Bragg spectroscopy technique to resolve its collective modes. We observe dissipation-induced synchronization at the quasiparticle level, where two roton-like modes coalesce at an exceptional point. This reveals how dissipation microscopically drives collective dynamics and signals a precursor to a dynamical phase transition. 
\end{abstract}

\maketitle

Synchronization -- a process where coupled oscillators adjust their frequencies to oscillate in unison -- is a universal phenomenon observed across physics, biology, sociology, and neuroscience~\cite{Pikovsky2001}. Classical examples range from synchronized pendulum clocks~\cite{huygens1916} to cardiac-respiratory coupling in humans~\cite{Holstege2014}, coordinated crowd clapping~\cite{neda2000}, and neuronal firing patterns~\cite{fries2015}. Synchronization requires sufficiently strong coupling between oscillators and system non-linearity, which can  be  induced via dissipation\,\cite{Lu2023Synchronization}. Growing interest in synchronization in quantum many-body systems\,\cite{Cabot2019Quantum,ValenciaTortora2023Crafting,Schmolke2024Measurement,poli2024sync} has revealed links to non-equilibrium dynamics\,\cite{gupta2014Kuramoto}, the interplay between entanglement and synchronization\,\cite{manzano2013sync,Witthaut2017Classical}, and the emergence of $PT$-symmetry breaking, limit cycles, and time-crystalline phases\,\cite{Fruchart2021Non,Nadolny2025Nonreciprocal}.

Transitions from stationary to non-stationary phases in such systems are particularly intriguing. In equilibrium, second-order phase transitions typically involve the softening of a collective excitation mode, such as roton softening preceding crystalline order \cite{nozieres2004}. Recent theoretical studies suggest that analogous behavior  occur in open quantum systems, where a collective response diverges at finite frequency, driving the dynamics of an emergent synchronized phase~\cite{Hanai2020Critical,Fruchart2021Non,Nie2023}. 
These transitions are non-reciprocal, since dissipation breaks the symmetry of mutual mode interactions and can take place at exceptional points (EP), which are spectral degeneracies unique to non-Hermitian systems, where both eigenvalues and eigenvectors coalesce\,\cite{Fruchart2021Non}.
Yet, a microscopic understanding of how synchronization arises from the interplay between dissipation, collectivity, and many-body excitations remains incomplete.

Ultracold atomic gases in optical cavities provide a unique platform for exploring non-equilibrium phases, as their interactions, geometry, and dissipation can be precisely controlled. Recent advancements in experimental techniques have enabled the realization of self-organized superradiant phases\,\cite{baumann2010,li2021}, arising from the softening of roton-like modes\,\cite{mottl2012}, as well as the observation of dynamical phenomena such as dynamical instabilities\,\cite{dogra2019}, an emergent topological pump\,\cite{dreon2022}, and time crystals\,\cite{Kongkhambut2022,Zhang2017, else2020discrete}.  

\begin{figure}[h]
\includegraphics[width=1\columnwidth]{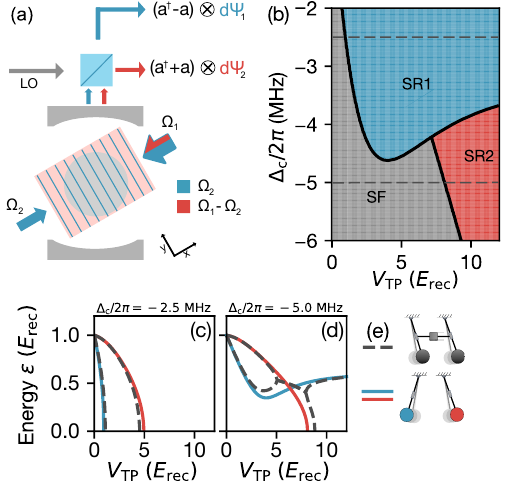}  
\caption{\textbf{Setup schematic with phase diagram and mode energies } \textbf{(a)} Sketch of the system's geometry. A BEC is coupled to a cavity and driven by an imbalanced repulsive field. The standing - and running-wave component of the drive couple to different quadratures of the cavity field. 
\textbf{(b)} Numerically calculated phase diagram as a function of $\Delta_c$ and $\VTP$ showing the two superradiant (SR) phases and the standard superfluid (SF), for an imbalance parameter of $\gamma=1.3$. 
\textbf{(c,d)}  Mode energies of the precursor mode for the SR1\,(blue) and SR2\,(red) phases as a function of the transverse pump, for $\Delta_c/2\pi=[-2.5(\text{c}); -6.0(\text{e})] \SI{}{\mega\hertz} $. With increased $\VTP$, the system experiences mode softening due to the proximity to a structural phase transition to either one of the two SR phases. The dark dashed lines show the induced synchronization effect at the mode crossing due to the finite cavity decay rate $\kappa$. (e) Mechanical analogy for dissipative synchronisation: two pendulums synchonize  when their relative motion is sufficiently damped. }
\label{fig:concept}
\end{figure}

Here, we experimentally realize synchronization and provide a microscopic characterization of the underlying mechanism in a many-body  cavity-QED system.
We introduce a  heterodyne-based Bragg spectroscopy scheme that  resolves coexisting collective excitation modes, the system's quasi-particle excitations, even in regimes where their spectral features overlap. This allows us to identify a transition to synchronization when two distinct collective modes become degenerate at finite energy and coalesce at an exceptional point. 
In this regime, dissipation synchronizes fluctuations of the bare modes, shaping the system's non-equilibrium dynamics.

Our setup consists of a quantum gas  characterized by long-range, cavity-mediated interactions that can induce atomic self-organization\,\cite{mivehvar2021cavity}, see Figure\,\ref{fig:concept}(a). Its phase diagram includes a normal phase and two distinct centro-symmetric, self-ordered crystals belonging each to a $Z_2$ symmetry class~\cite{li2021}. The emergence of these two crystalline phases is  driven by the softening of two distinct collective excitation modes. Photon leakage from the cavity creates a non-Hermitian many-body setting, and leads to a dissipative coupling between these modes.  We  identify the modes, analyze the coupling,  and track their frequency evolution across the phase diagram, with a specific focus on their synchronisation.

Our results are supported by two  theoretical approaches: a mean-field framework based on the system’s band structure (detailed in  companion paper \cite{suppmat2}) and Gross-Pitaevskii equation (GPE) simulations. Together, they establish a microscopic understanding of synchronization as a collective phenomenon among quasi-particle excitations, revealing how multiple softening modes can lock to a common finite frequency and act as a precursor to a dynamical phase transition\,\cite{dreon2022}.

 We produce a Bose-Einstein condensate (BEC) of \mbox{$N=2.5(3)\cdot10^5$} $^{87}$Rb atoms in an optical dipole trap positioned at the center of a high-finesse optical cavity. The cavity has a single-atom vacuum Rabi coupling rate $g/2\pi = \SI{1.95}{\mega\hertz}$ and a field decay rate $\kappa/2\pi = 147$~kHz, providing the dissipation\,\cite{leonard2017supersolid,zupancic2019,dreon2022}. We employ two imbalanced counterpropagating pump beams arising from a single laser field blue-detuned from the atomic transitions, with frequency $\omega_\text{p}$, incident transversely at an angle of 60° relative to the cavity axis\,\cite{li2021}. The two counterpropagating beams travel along the $x$ direction and have coupling strenghts $\Omega_1$ and $\Omega_2$, whose tunability allow us to vary the imbalance parameter $\gamma=\sqrt{\Omega_2 / \Omega_1}$. 
The blue-detuned and imbalanced character of the transverse pump, with lattice depth $\VTP$ leads to two distinct self-ordered superradiant phases: blue detuning pushes atoms into intensity minima, while imbalance maintains a nonzero field there, enabling the emergence of two distinct phases~\cite{li2021}. The detuning $\Delta_c=\omega_\text{p} - \omega_\text{c}$ between the pump laser and the cavity resonance $\omega_\text{c}$ together with $\VTP$ serve as tunable parameters that control the onset of these superradiant phases. By using a heterodyne detection scheme\,\cite{li2021}, we record the light field leaking out of the cavity and extract the field amplitude $\alpha$ and phase $\phi$, allowing a continuous non-destructive probe of the system.

An intuitive understanding of the mode softening responsible for the two superradiant phases is obtained from a 2D model with $\kappa=0$, based on a mean-field approach and a few-mode ansatz for the atomic wavefunction, see  \,\cite{suppmat2}. Fig.\,\ref{fig:concept}(b) shows the resulting phase diagram as a function of $\Delta_\text{c}$ and $V_\text{TP}$. We identify three phases: the superfluid BEC in the optical lattice provided by the pump field (SF),  and two superradiant phases (SR1, SR2).  Phase SR1 arises from coupling to the standing-wave component of the drive and is present even in the balanced pump configuration, whereas SR2 originates from the running-wave component introduced by the pump imbalance\,\cite{li2021}. The two superradiant phases exhibit distinct symmetries and couple to spatial modes of opposite parity, $\cos(\mathbf{k}_\text{c} \cdot \mathbf{r})$
and $\sin(\mathbf{k}_\text{c} \cdot \mathbf{r})$, where $\mathbf{k}_\text{c}$ is the cavity wave vector~\cite{li2021, dreon2022}. This leads to scattering into orthogonal light quadratures $P = \frac{i}{\sqrt{2}} \langle \hat{a}^\dagger - \hat{a} \rangle$ and $Q= \frac{1}{\sqrt{2}} \langle \hat{a}^\dagger + \hat{a} \rangle$. Here, $\hat{a}$ is the bosonic annihilation operator for a cavity photon. Each phase features a soft excitation mode that vanishes at the boundary to the SF phase. The meeting point of the boundaries between the SF, SR1, and SR2 phases suggests the presence of a tricritical point, where the normal phase becomes simultaneously unstable to both superradiant orders, implying that the precursor modes of SR1 and SR2 soften together at this point.

\begin{figure*}[ht]
\centering
\includegraphics[width=1\textwidth]{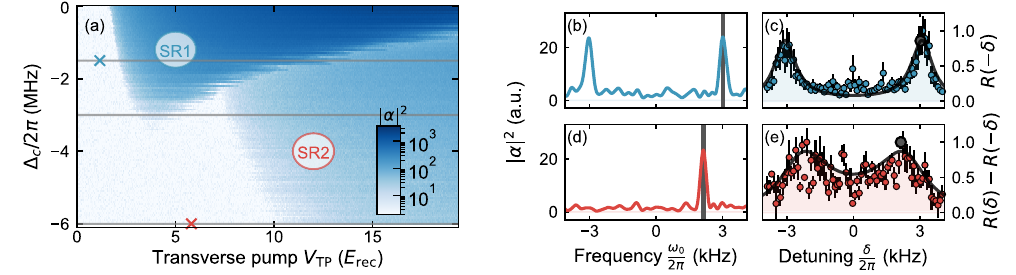}
\caption{\textbf{Measured phase diagram and Bragg responses} \textbf{(a)} Phase diagram for an imbalanced pump as a function of  $\Delta_c$ and  $\VTP$. Color shading indicates the intensity of the cavity field $|\alpha|^2$. Three phases are present: The superfluid phase (without cavity field, white area) and two different superradiant phases (SR1, SR2). The grey lines indicate the cavity detunings explored in the Bragg spectroscopy measurements shown in Fig.\,\ref{fig3:Double-roton},\,\ref{fig5:Syncronization}
\textbf{(b,d)} Cavity field spectra after probing the system close the to SR1 phase (b) ($\Delta_c = 2\pi \times \SI{-1.5}{\mega\hertz}$, $\VTP$ = \SI{1.16(0.04)}{\Er}, blue cross in (a)) and close to the SR2 phase (d) ($\Delta_c = 2\pi \times \SI{-6}{\mega\hertz}$, $\VTP$ = \SI{5.8(2)}{\Er}, red cross in (a)). Close to the SR1 phase, the response occurs at both the probed frequency (gray line in the plot) and its negative counterpart, while close to  the SR2 phase, the response is only at the probed frequency.
\textbf{(c,e)} Response $R(-\delta)$ (c) and $R(\delta)-R(-\delta)$ (e) as a function of the detuning $\delta$ between probe and pump frequencies for the SR1 mode (same parameters as (b)) and the SR2 mode (same parameters as (d)), respectively. The enlarged gray points highlight the detunings $\delta$ chosen for the on-resonance responses in (b,d).  The error bars represent the standard error of the mean for typically 5 repetitions. The black line indicates the fit to the data, see\,\cite{suppmat}.}
\label{fig:PD}
\end{figure*}

Figures~\ref{fig:concept}\,(c-d) show the calculated  energies of the SR1 and the SR2 modes as function of the transverse pump strength for two different $\Delta_\text{c}$. At $\VTP=0$, both modes start at $\SI{1}{\Er}$, the recoil energy. Increasing $\VTP$ initially reduces both mode energies, but their behavior depends on the detuning. For small detuning (panel~c), both modes soften completely, with SR1 softening faster than SR2.
At larger detuning (panel~d), the SR1 mode softens and rehardens beyond the tip of the SR1 phase in the phase diagram (see panel b) at $\VTP \approx \SI{4}{\Er}$, while the SR2 mode fully softens at higher $\VTP$, resulting in a mode crossing of the SR1 and the SR2 modes. 
At this intersection, the energy gap between the modes vanishes, which will, in presence of dissipation, lead to coupling between the two modes and favor their synchronization. To provide an intuitive picture, we draw an analogy to two pendulums coupled by a damping element (panel~e), which synchronize when the dissipation is sufficiently strong. In our system, the pendulums correspond to fluctuations of two distinct quasiparticle excitations, and the damping arises from the coupling of the cavity field to the electromagnetic vacuum. This simplistic analogy illustrates how dissipation can mediate synchronization between modes.

We experimentally map out the phase diagram as a function of detuning $\Delta_c$ and pump lattice depth $\VTP$, see Fig.~\ref{fig:PD}(a) for a fixed imbalance parameter $\gamma=1.4(1)$. We linearly ramp $\VTP$ from 0 to \SI{19.4(7)}{\Er} within \SI{50}{\milli\second}, repeating the same experimental sequence for various values of $\Delta_c$. The ramping time is chosen to ensure adiabaticity with respect to the trapping frequencies, while remaining short enough to keep atom loss low~\cite{suppmat}. We continuously record the light field leaking out of the cavity and extract the field amplitude $\alpha$ as function of $\VTP$ and $\Delta_c$. The resulting phase diagram reveals three distinct phases as calculated by our model. In the superfluid phase, we detect only the vacuum noise of the cavity mode. In the two self-organized phases, SR1 and SR2, nonzero average intracavity fields are observed.

To obtain the energies of the SR1 and SR2 modes, we perform cavity-assisted Bragg spectroscopy: After linearly increasing $\VTP$ from 0 to a final value in \SI{50}{\milli\second}, we inject a coherent probe field into the cavity\,\cite{mottl2012}.
The interference between the probe field and the transverse pump creates a grating potential on the atomic cloud. The probe field frequency $\omega_{\text{probe}}$ is detuned with respect to the transverse pump by a variable frequency  $\delta = \omega_{\text{probe}} - \omega_\text{p}$. Due to the imbalance in the transverse pump, the grating potential leads to an amplitude as well as a phase modulation,  providing a probe for each of the two roton-like modes\,\cite{suppmat}.

We present the response of the system for two distinct cavity detunings and $\VTP$ : \(\Delta_c = 2\pi \times \)\SI{-1.5}{\mega\hertz}, $\VTP=\SI{1.16(0.04)}{\Er}$ (Fig.\,\ref{fig:PD}(b,~c)) and \(\Delta_c =  2\pi \times \)\SI{-6}{\mega\hertz}, $\VTP=\SI{5.8(2)}{\Er}$ (Fig.\,\ref{fig:PD}(d,~e)). Figures \,\ref{fig:PD}(b,~d) display the frequency spectrum of the cavity field, measured via heterodyne detection, when the detuning $\delta$ matches the respective roton modes energies. For the SR1 mode (Fig.\,\ref{fig:PD}(b), we measure a  symmetric response, appearing at $\pm \omega_{\text{probe}}$, while for the SR2 mode (Fig.\,\ref{fig:PD}(d)) the response is observed to be asymmetric, only at $\omega_{\text{probe}}$.  In the first case, the perturbation predominantly excites the SR1 mode; in the second case, SR2.

Motivated by this observation, we analyze two complementary observables: $R(-\delta)$ and $R(\delta) - R(-\delta)$, where $R(\delta)$  is the normalized response extracted from the frequency spectra integrated over a frequency interval centered at $\delta$ and width set by the Fourier broadening. These observables not only enable accurate determination of the central frequency of the modes, but also allow isolating contributions from the two superradiant phases due to their different symmetries. Figure \,\ref{fig:PD}(c,~e) show the response $R(-\delta)$ and $R(\delta) - R(-\delta)$, respectively. 
The resonances appear at the energy of the respective roton mode, displaying a spectral profile determined by the interference between the injected probe field and the Bragg-scattered pump field\,\cite{mottl2012,suppmat}.

\begin{figure}[t]
\centering
\includegraphics[width=1\columnwidth]{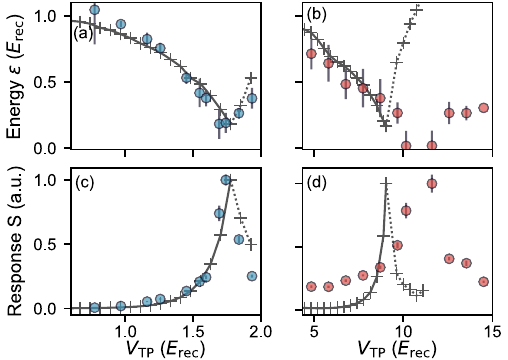}
\caption{\textbf{Softening SR1 and SR2 mode energies and divergent response amplitudes.} \textbf{(a) [(b)]} Measured softening of the SR1 [SR2] mode energy $\epsilon$ as a function of pump strength $\VTP$ at  $\Delta_c = 2\pi \times -1.5$ \si{\mega\hertz} [$\Delta_c = 2\pi \times  -6$ \si{\mega\hertz}]. Error bars represent $68\%$ confidence intervals, see\,\cite{suppmat}.
\textbf{(c, d)} Normalized response $S$ as a function of $\VTP$ showing the corresponding divergence of the susceptibility for the parameters in (a,b). Error bars represent the fit's statistical uncertainty.  The gray lines represent the  result from the GPE model. The $\VTP$-axes for the numerical results have been rescaled by a constant factor $c=1.35$ . Simulations for the ordered phase are shown as dotted line.}
\label{fig3:Double-roton}
\end{figure}

From the response functions, we extract the central frequency, i.e. the mode energy \(\epsilon\), and the response amplitude $S$ by fitting the data to a model \cite{suppmat,mottl2012}, see Fig.\ref{fig3:Double-roton}. For \(\Delta_c = 2\pi \times -1.5\)~\SI{}{\mega\hertz} (a, c), the SR1 mode exhibits a pronounced softening with increasing  $\VTP$, as seen in the gradual reduction of mode energy \(\epsilon\) towards the critical point. Concurrently, the response \(S\) reveals a sharp increase consistent with the divergence of susceptibility at a phase transition.
At \(\Delta_c = 2\pi \times -6\)~\SI{}{\mega\hertz} (b, d), a similar trend is observed for the SR2 mode, albeit at higher $\VTP$ compared to the SR1 phase. The delayed onset of the SR2 phase reflects the different critical conditions imposed by the cavity detuning and the imbalance parameter\,\cite{li2021}. 

To obtain a more complete description of the system response, we numerically solve a GPE which, different from the previously described simplified 2D model, also takes into account the trapping potential and the contact interactions\,\cite{suppmat,GPE_SciPostPhysCodeb}.
The GPE simulations show quantitative agreement with the experimental behavior of the SR1 mode, see panels (a) and (c). In panels (b) and (d), corresponding to the SR2 mode, the agreement is only qualitative, capturing the overall softening trend and response growth. To account for atom losses during the experiment, the GPE results are rescaled by a global factor along the transverse-pump axis. We assign the remaining discrepancies between GPE theory and experiment to non-adiabaticity and to the pump-strength-dependent losses~\cite{baumgartner2024stability} not captured by our global rescaling\,\cite{suppmat}.

\begin{figure}[t]
\centering
\includegraphics[width=1\columnwidth]{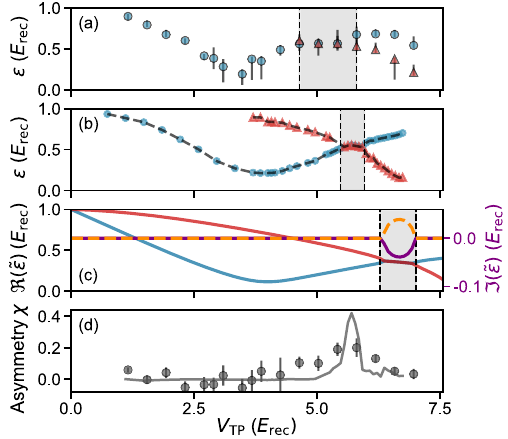}
\caption{\textbf{Mode Synchronization at $\Delta_c =2\pi \times  -3$ \si{\mega\hertz}. } \textbf{(a)} Experimentally measured mode energies $\epsilon$ as a function of  transverse pump strength $\VTP$. Data points in blue(red) refer to the SR1(SR2) mode.  The error bars represent $68\%$ confidence intervals, see\,\cite{suppmat}.\textbf{(b)} Mode energies extracted from  GPE simulations in presence of dissipation. The dashed line  is  a guide to the eye. The $\VTP$-axis is rescaled by a constant factor $c=1.35$. \textbf{(c)} Real (blue, red) and imaginary (orange, purple)  parts of the eigenvalues from the  band-structure model at $\Delta_c = 2\pi \times -4.5,\si{\mega\hertz}$, capturing the synchronization.   \textbf{(d)} Asymmetry $\chi$ in the spectral response as a function of $\VTP$. The error bars represent the standard error. The solid line indicates  the GPE simulations.}
\label{fig5:Syncronization}
\end{figure}

We now turn to an intermediate cavity detuning of $\Delta_c = 2\pi\times \SI{-3}{\mega\hertz}$, where the two excitation modes are expected to cross outside the superradiant phases. Figure~\ref{fig5:Syncronization}(a) displays the experimental data as a function of the transverse pump strength: At low $V_{\text{TP}}$, the SR1 mode undergoes a partial softening, before it starts to harden again around $3.5~E_\text{rec}$.  For small $V_{\text{TP}}$, the SR2 mode remains undetectable due to the dominant response of the SR1 mode~\cite{suppmat}. The SR2 mode becomes visible only at higher $V_{\text{TP}}$, as the system approaches the critical point of the SR2 phase. In the intermediate regime, we observe the measured mode energies overlap between $4.6(2)~E_\text{rec}$ and $5.8(3)~E_\text{rec}$ in the experimental data, signaling their synchronization. Figures~\ref{fig5:Syncronization}(b) and (c) present numerical results from the GPE and the 2D analytical model, respectively. The GPE simulation predicts synchronization between $5.5$ and $6.0~E_\text{rec}$, while the simplified band structure model, which neglects interactions and external confinement, yields a synchronization region between $6.3$ and $7.0~E_\text{rec}$. The deviations reflect the same limitations discussed for Fig.\,\ref{fig3:Double-roton}, see\,\cite{suppmat}.

Dissipation plays a crucial role in this process; when omitted from simulations, synchronization does not occur and the modes just cross (see Fig.~\ref{fig:concept}(d,e)). From the simplified model, we extract  the eigenvalues and the eigenvectors, corresponding to the complex energies ($\tilde{\epsilon}$) and wavefunctions of the two modes~\cite{suppmat2}. Figure\,\ref{fig5:Syncronization}(c) illustrates the evolution of the real and imaginary parts of the eigenvalues. At the two specific pump strengths where the eigenvalues coalesce, the eigenmodes also coincide, signaling the presence of  EP.  Between these two EPs, the real parts of the eigenvalues remain identical, while their imaginary parts differ.  One hybridized mode acquires a positive imaginary component, indicating gain, and the other a negative component, signifying loss if the system evolves in time. In this regime, the two eigenvectors do not correspond to the initial mode wavefunctions, because the dissipative coupling induces a distinctive mode mixing\,\cite{suppmat2}. This mixing leads to a dynamical phase in which the cavity field oscillates with the real eigenfrequency of the hybridized mode between the P and Q quadratures of the cavity field. The selection of a rotation direction is a hallmark of the \textit{PT}-broken phase that emerges between two EPs.
 
Experimentally, we observe this chirality through the asymmetry parameter  $\chi = \frac{\tilde{R}_{\text{max}}^+ - \tilde{R}_{\text{max}}^-}{\tilde{R}_{\text{max}}^+ + \tilde{R}_{\text{max}}^-}$, which quantifies the imbalance in the cavity field response. The quantities $\tilde{R}_{\text{max}}^{\pm}$ represent the maximum integrated response amplitudes at positive and negative probe detunings, respectively,\,see \cite{suppmat}. A positive (negative) value of $\chi$ indicates a clockwise (counterclockwise) rotation in the P–Q quadrature space. This behavior is illustrated in Fig.\,\ref{fig5:Syncronization}(d), where the measured asymmetry of the response in the synchronized regime agrees qualitatively with our  results from the  GPE simulation. Synchronization in our system is identified directly through two independent observables: the merging of the mode energies and the appearance of chirality from the asymmetry parameter $\chi$. The driven-dissipative many-body system supports collective quasi-particle excitations that  synchronize under suitable conditions. This synchronized behavior closely resembles that of two harmonic oscillators coupled via dissipation, whose resonance frequencies intersect~\cite{Lu2023Synchronization}, offering an intuitive analogy for understanding the underlying dynamics. Within a certain range of frequency mismatch around this intersection, the two oscillators become phase-locked and evolve at the same frequency.

In conclusion, we developed a Bragg spectroscopy technique to resolve two quasi-particle modes associated with competing superradiant phases in a BEC coupled to an optical cavity. As the transverse pump strength increases, the modes soften and eventually synchronize near a tricritical point, driven by the interplay of long-range interactions and cavity-mediated dissipation. This synchronization is marked by the emergence of EP and is supported by two  theoretical models. Our findings provide a new perspective on the self-driven topological pump, going beyond the observation in Ref.\,\cite{dreon2022}; by probing the system spectroscopically from the normal phase, we uncover mode synchronization as the underlying mechanism driving the phenomenon.

Our results demonstrate the power of mode-resolved spectroscopy to uncover the microscopic dynamics shaping non-equilibrium phases in open quantum systems. They also reveal a fundamental link between equilibrium second-order phase transitions, where a single mode softens to zero energy, and the synchronization phenomenon observed here,  giving rise to a so-called critical exceptional point~\cite{Hanai2020Critical}. While both scenarios exhibit criticality, they emerge from fundamentally different mechanisms. Theoretical work suggests that such non-Hermitian, synchronization-driven transitions belong to distinct dynamical universality classes~\cite{Sieberer2013Dynamical} and are a hallmark of non-reciprocal phase transitions shaped by dissipation and directional coupling~\cite{Fruchart2021Non}.

Future experiments will probe  critical scaling near these transitions, offering insight into a broader class of driven systems -- from ultracold quantum gases to biological and artificial neural networks.  Ultracold atoms thus provide a uniquely controlled setting to probe the universal features of non-reciprocal matter, characterized by interactions that lack reciprocity. 

\hfill \break

\textit{Acknowledgements---}
We thank Alexander Frank for assistance with the electronic equipment. This research was supported by the Swiss National Science Foundation (SNSF) project numbers\,186312, 212168, 217124, 221538, 223274, and the Swiss State Secretariat for Education, Research and Innovation (SERI) under grant number\,MB22.00090. This project is funded within the QuantERA II Programme, which has received funding from the EU's Horizon 2020 research and innovation programme under Grant Agreement No. 101017733.
D.R.  acknowledges support from the Brazilian funding agency Conselho Nacional de Desenvolvimento Científico e Tecnológico (CNPq Process No. 201090/2022-8 and  CNPq Process No. 402660/2019-6).

\textit{Data availability---}
The data that support the findings of this article are openly available\,\cite{RCsync} 
\bibliography{bibliography}

\newpage
\begin{widetext}
\begin{appendix}

\section{Supplementary material}

\subsection*{Production of the Bose-Einstein condensate (BEC)}
The experimental procedure begins with creating a $^{87}$Rb Bose-Einstein condensate (BEC) containing $N=2.5(3)\times 10^5$ atoms at a temperature of $102(10)\mathrm{nK}$. 
The BEC is confined in an optical dipole trap, with trapping frequencies of $[\omega_\textrm{x}, \omega_\textrm{y}, \omega_\textrm{z}]=2\pi \times [89(1), 70(3), 219(3)] \mathrm{Hz}$, which is used to position the BEC at the center of the cavity mode.

The atomic cloud is subject to losses due to heating or off-resonant scattering. The loss rate depends on the applied transverse pump lattice depth and whether the system is self-ordered, with typical loss rates between 2~Hz and 5~Hz~\cite{baumgartner2024stability}.

The coupling between the atoms and the cavity is proportional to the overlap between the cavity mode and the atomic cloud. 
This coupling can be assessed by measuring the dispersive shift in the cavity resonance. 
During the measurement, a magnetic offset field of approximately $25\mathrm{G}$ is applied to prevent spin-dependent effects on the self-organization, such as superradiance into the orthogonally polarized cavity mode. 
This mode is significantly detuned due to the birefringence of our cavity, as previously studied in\,\cite{morales2019}.

\subsection*{Cavity parameters and phase diagram for our system}
The cavity exhibits a decay rate of $\kappa = 2\pi \times 147(4)\,\mathrm{kHz}$ and has a maximum single-atom vacuum Rabi frequency of $g = 2\pi \times 1.95(1)\,\SI{}{\mega\hertz}$. The cavity mode is oriented at an angle of $60^{\circ}$ relative to the direction of the pump light field. This pump light is partially retroreflected onto the atoms, creating a standing wave lattice. The part of the pump light not compensated by the retroreflected light results in a running wave potential. By adjusting the fraction of the reflected light, we control the relative contribution of the standing and running wave components.
The pump beam frequency $\omega_\textrm{p}$ is blue-detuned from the $\text{D}_2$-line of $^{87}$Rb, $\omega_\textrm{a}$, by $\Delta_\textrm{a}=\omega_\mathrm{p}-\omega_\textrm{a}=2\pi \times 69.8(1)\,\mathrm{GHz}$.
The relative detunings between the cavity resonance frequency $\omega_\textrm{c}$ and the transverse pump, $\Delta_\textrm{c}=\omega_\mathrm{p}-\omega_\textrm{c}$, can be tuned using an electro-optical modulator. 
The cavity length is stabilized with low-amplitude, far-detuned laser field at $830\,\mathrm{nm}$, used to provide continuous feedback on the cavity lengths while minimally affecting the atomic cloud.

During each experimental cycle, the transverse pump lattice depth $\VTP$ is linearly increased over $50\,\mathrm{ms}$ from $0$ to a maximum of $19.4(7)\,\mathrm{E_\textrm{rec}}$, where the recoil energy $E_\textrm{rec}=2\pi\hbar \times 3.77\,\mathrm{kHz}$. 
Photons leaking from the cavity are measured and used to determine the intracavity photon number. 
For certain cavity detunings and beyond a critical pump strength, the intracavity field becomes populated as the atoms self-organize in a density pattern, which can scatter photons from the transverse pump into the cavity. 
We repeat the transverse pump ramp at various cavity detunings and generate the phase diagram, as illustrated in Fig.\,\ref{fig:PD}a.
The calibration of the transverse pump lattice depth and cavity lattice depth was performed using Raman-Nath diffraction.

\subsection*{Heterodyne Technique}

The amplitudes of the intracavity fields are measured using balanced optical heterodyne detectors. We use a local oscillator (LO) light field as a reference for measuring the phase shifts due to the physical processes in our experiment.
The LO frequency is offset from the transverse pump laser source by $50\,\text{MHz}$ using an acousto-optic modulator. We combine the LO with the leaked light from the cavity at a beam splitter, directing the resulting light onto the two photodiodes of a balanced photodetector. 
The obtained signal is downmixed to a radio frequency of $300\,\text{kHz}$. Signal data at $300\,\text{kHz}$ is captured via a computer-connected oscilloscope (PicoScope 5444b). Subsequently,  we extract the photon field by  applying a low-pass filter using a binning window of $1\times10^{-4}$ s. All radio frequency sources are phase-locked to a $10\,\text{MHz}$ GPS frequency standard with a fractional stability exceeding $10^{-12}$ to minimize technical phase noise in the heterodyne detection setup. For detailed technical aspects and design considerations, we refer to\,\cite{li2021}. 
\subsection*{Protocol for Bragg Spectroscopy: Methodology and Analysis}

Since we are primarily interested in studying the mode structure and its softening, most measurements are performed before the system enters a superradiant phase. 
This corresponds to the superfluid (SF) region of the phase diagram. 
In a typical Bragg measurement sequence, after the initial transverse pump ramp to the desired $\VTP$, we hold the transverse pump depth constant and apply a weak on-axis cavity probe. 
This probe, with a constant amplitude, acts as a drive to the cavity field. 
The probe frequency, \(\omega_{\text{probe}}\), is detuned by \(\delta = \omega_{\text{probe}} - \omega_{\text{p}}\) from the pump frequency. 
Consequently, the interference between this beam and the transverse pump creates a modulated potential acting as a Bragg grating, enabling spectroscopic measurements \cite{mottl2012}.
We detect photons leaking from the cavity using a heterodyne setup, which provides real-time access to the amplitude and frequency of the cavity light. 
Due to the non-destructive nature of our measurements, we perform Bragg spectroscopy for an arbitrarily long pulse\,(\SI{10}{\milli\second}) and fix \textit{a posteriori} the analyzed time. For this study, we choose a $\tau=\SI{3}{\milli\second}$  evaluation time to minimize the linewidth of the system's response. Shorter pulse durations are limited by Fourier broadening, whereas longer pulses yield no further reduction in linewidth, suggesting a finite mode lifetime.
By taking the Fourier transform of the $3$\,ms probing window, we can analyze the spectral components of the detected signal. For a given probe frequency, we extract two observables: the response $R(\delta)$, which measures the number of cavity photons at the probe frequency $\delta$, and $R(-\delta)$, which measures the number of cavity photons at $-\delta$ for the same probing frequency $\delta$. More specifically we define the response functions as:
\begin{equation}
R(\delta) =\int_{\delta - \Delta f}^{\delta + \Delta f}\mathrm{d}f \, \abs{\alpha(f)}^2, \quad R(-\delta) = \int_{-\delta - \Delta f}^{-\delta + \Delta f} \mathrm{d}f \, \abs{\alpha(f)}^2,
\label{eq:SuppResponseR}
\end{equation}
where \( \Delta f = 300 \,\text{kHz} \) is set by the Fourier broadening of the probe pulse.
By varying \(\delta\) over \(\pm 4\) kHz, we study how the response changes. 
In the response data of $R(\delta)$, and $R(-\delta)$, we observe a clear resonance behavior, which we analyze following reference\, \cite{mottl2012}. In brief, the intracavity photon number ($n_\mathrm{ph}$) during the Bragg pulse is due to three terms. The first is the coherent cavity amplitude $\alpha_0$ present only in the superradiant phases, the second is the transverse pump being Bragg-scattered in the cavity due to the excited population and the last term is the probe field externally imposed $n_\mathrm{pr}$:
\begin{equation}
n_{\mathrm{ph}}(t)=\left\lvert\, \alpha_0-\frac{A}{\tilde{\Delta}_{\mathrm{c}}+i \kappa} \operatorname{Im}(\mathcal{Y}(t))\right.  +\left.\sqrt{n_{\mathrm{pr}}(t)} e^{-i(\delta t+\varphi)}\right|^2,
\label{eq:Si_Nphoton}
\end{equation}
here $\mathcal{Y}$ is: 
\begin{equation}
\mathcal{Y}(t)=e^{\left(i \epsilon-\gamma\right) t} \int_0^t \mathrm{~d} t^{\prime} e^{-\left(i \epsilon-\gamma\right) t^{\prime}} \cos \left(\delta t^{\prime}+\varphi\right) \Pi\left(t^{\prime}\right),
\end{equation}
The relative phase between the probe field and the transverse pump $\phi$, is
not controlled in the experiment, and varies between different experimental runs. 
To apply this fitting model to our heterodyne data we calculate $R(\delta)$ and $R(-\delta)$ from the fourier transform of eq.\,\eqref{eq:Si_Nphoton} using \eqref{eq:SuppResponseR}.
In the final expression we have three fitting parameters $A$, $\gamma$ and $\omega_s$.
The first term $A$ depends on the atom number, the strength of the probing perturbation and the susceptibility.  The second term $\gamma$ is a damping rate due to a finite lifetime of the excitation. Finally the last term $\epsilon$ indicate the energy of the soft mode.

To study mode softening, we repeat these measurements for different transverse pump strengths and plot the resonance frequency ($\epsilon$) and resonance amplitude (S) as a function of the pump strength. 
For different transverse pump strength, the system response S can be obtained after rescaling A by the square root of the transverse pump depth, see later discussion.
Our heterodyne detection scheme allows for the simultaneous measurement of $R(\delta)$ and $R(-\delta)$ for a single probe frequency $\delta$, enabling us to distinguish the precursor mode responses associated with the two superradiant phases. 
The softening modes corresponding to the two superradiant phases show a distinct response. The mode corresponding to the phase 1 (SR1) shows an stronger amplitude response, with a response at both $R(\delta)$ and $R(-\delta)$, while the mode corresponding to phase 2 (SR2) primarily shows a response at $R(\delta)$. 
Consequently, we define independent observables for each mode and analyze the softening of the SR1 with \(R(-\delta)\) and SR2 using the difference $R(\delta)-R(-\delta)$.
This response function is shown in Fig.\,\ref{fig:PD}, where we compare two different cavity detunings  $\Delta_\textrm{c}=2\pi \times  \,\SI{-1.5}{\mega\hertz}$ and $\Delta_\textrm{c}=2 \pi \times-\,\SI{6.0}{\mega\hertz}$.

To estimate the confidence interval of the energy of the soft mode, we use likelihood profiling.  After determining the optimal parameter values, we scan the parameter of interest over a sufficiently broad range, fixing it at each value while reoptimizing the remaining model parameters. The resulting change in the cost function ($\chi^2$) is then computed.  The confidence interval is then defined as the range of parameter values for which the increase in $\chi^2$ remains within the threshold corresponding to $68\%$ confidence. This approach may result in asymmetric confidence intervals.

\subsection*{Extraction of the Spectral Asymmetry}

To characterize the synchronized regime and probe the directional response of the soft modes, we extract a spectral asymmetry parameter \( \chi \) from the Bragg spectroscopy data. This parameter quantifies the imbalance between the cavity responses at positive and negative probe detunings.
For each probe detuning \( \delta \), we compute the integrated response around \( \omega = \pm \delta \) by summing the detected signal over a fixed frequency window of \(10\,\text{kHz} \):
\begin{equation}
\tilde{R}(\pm\delta) = \int_{- 5\,\text{kHz}}^{ + 5\,\text{kHz}} \mathrm{d}f \, \abs{\alpha(f)}^2.
\end{equation}
The integration range of \SI{\pm 5}{\kilo\hertz} around the probe frequency is chosen to encompass the full spectral response of the system, as the signal is well-contained within this window. Extending the integration limits further would not significantly change the extracted signal but would increase the contribution of technical noise. Thus, the range serves as a practical approximation to an infinite integration domain while effectively suppressing noise contributions.
We then extract the maximum response across all scanned detunings:
\begin{equation}
\tilde{R}_{\text{max}}^{+} = \max_{\delta > 0} \left(\tilde{R}(\delta)\right), \quad \tilde{R}_{\text{max}}^{-} = \max_{\delta > 0} \left(\tilde{R}(-\delta)\right).
\end{equation}
The asymmetry parameter \( \chi \) is defined as:
\begin{equation}
\chi = \frac{\tilde{R}_{\text{max}}^{+} - \tilde{R}_{\text{max}}^{-}}{\tilde{R}_{\text{max}}^{+} + \tilde{R}_{\text{max}}^{-}}.
\end{equation}
This measure allows us to track the emergence of a preferred frequency sign in the synchronized regime. As shown in Fig.\,\ref{fig5:Syncronization}(g), the asymmetry \( \chi \) displays a resonant behavior centered around the critical transverse pump strength where the two roton-like modes coalesce. This observation matches our theoretical prediction and signals the onset of dissipatively induced mode synchronization.

\section{Hamiltonian of the system}

The full single-particle Hamiltonian describing the system reads:
\begin{align}
    \mathcal{\hat{H}}/\hbar =& \frac{\hat{\mathbf{p}}^2}{2m} +  \VTP \cos^2(\mathbf{k_p}\cdot \hat{\mathbf{x}}) - \Delta_\mathrm{c} \hat{a}^\dagger \hat{a} + U_0\hat{a}^\dagger \hat{a}\cos^2(\mathbf{k_c} \cdot\hat{\mathbf{x}}) \nonumber\\
    &+ \sqrt{\VTP U_0}\left(\gamma+\frac{1}{\gamma}\right) \frac{\hat{a}+\hat{a}^\dagger}{2}\cos(\mathbf{k_p} \cdot\hat{\mathbf{x}})\cos(\mathbf{k_c} \cdot\hat{\mathbf{x}})\nonumber \\ 
    &+ \sqrt{\VTP U_0}\left(\gamma-\frac{1}{\gamma}\right) \frac{\hat{a}-\hat{a}^\dagger}{2i}\sin(\mathbf{k_p} \cdot\hat{\mathbf{x}})\cos(\mathbf{k_c} \cdot\hat{\mathbf{x}}),
    \label{eq:appendix:Hamiltonian:main}
\end{align}
where $\hat{\mathbf{p}}$ and $\hat{\mathbf{x}}$ are the single particle momentum and position operators. The cavity photons are described by the bosonic creation and destruction operators $\hat{a}^\dagger$ and $\hat{a}$. The strength of the transverse pump is given by $\VTP=\Omega_1\Omega_2/\Delta_a$. The coupling of the cavity light is determined by $U_0=g^2/\Delta_a$. Lastly, the imbalance between the two transverse pump beams is given by $\gamma=\sqrt{\Omega_1/\Omega_2}$.\newline
The two transverse pump beams lead to a standing wave lattice described by the second term along $\mathbf{k_p}$ and a constant offset, the latter is not taken into account in the following description. The cavity light results in a second standing lattice described by the fourth term and extends along $\mathbf{k_c}$. The interference of transverse pump and cavity light leads to two interference terms given in line two and three of Eq.~\eqref{eq:appendix:Hamiltonian:main}. The term on the second line is caused by the balanced part of the transverse pump, while the term on the third line is proportional to the imbalance between the two beams. Notably, the two terms couple to different quadratures of the light and have a different spatial pattern.\newline

To investigate the potential emerging from the Bragg-probe light in the cavity we consider the following two light fields:
\begin{align}
    t(x,t)&= (\Omega_1 e^{i\mathbf{k_p}\cdot \mathbf{x}}+\Omega_2 e^{-i\mathbf{k_p}\cdot \mathbf{x}} )e^{-i\omega_{p} t} + h.c.\\
    p(x,t)&= \Omega_p(\alpha e^{-i\omega_{\text{probe}} t}+\alpha^*e^{i\omega_{\text{probe}} t}) \cos(\mathbf{k_c} \cdot\mathbf{x}),
\end{align}
where $t(x,t)$ is the imbalanced transverse pump light field and $p(x,t)$ the probe light field in the cavity. If we calculate the absolute value of these two fields, we get the two familiar standing lattices along the transverse pump and cavity, as well as two interference terms.

\begin{align}
    \abs{t(x,t)+p(x,t)}^2=&2(\Omega_1-\Omega_2)^2+8\Omega_1\Omega_2\cos^2(\mathbf{k_p} \cdot\mathbf{x})+2\abs{\alpha}^2\Omega_p^2\cos^2(\mathbf{k_c} \cdot\mathbf{x})\nonumber\\
    &+4\Omega_p\abs{\alpha}(\Omega_1+\Omega_2)\cos(\mathbf{k_p} \cdot\mathbf{x})\cos(\mathbf{k_c} \cdot\mathbf{x})\cos(\Delta\omega t+ \phi)\nonumber\\
    &+4\Omega_p\abs{\alpha}(\Omega_1-\Omega_2)\sin(\mathbf{k_p} \cdot\mathbf{x})\cos(\mathbf{k_c} \cdot\mathbf{x})\sin(\Delta\omega t+ \phi),
\end{align}
where we have used $\alpha=\abs{\alpha}e^{i\phi}$.
Note, that due to the detuning of the cavity light field from the transverse pump both interference terms are amplitude modulated with the frequency difference: $\Delta\omega=\omega_p-\omega_{\text{probe}}$. The mode corresponding to SR1 responds to the term proportional to $\cos(\mathbf{k_p} \cdot\mathbf{x})\cos(\mathbf{k_c} \cdot\mathbf{x})$ whereas the SR2 precursor mode responds to the interference potential with the $\sin(\mathbf{k_p} \cdot\mathbf{x})\cos(\mathbf{k_c} \cdot\mathbf{x})$ spatial modulation.

\section{GPE Simulation}

The numerical simulations were performed by solving the Gross-Pitaevski equation, adapted to describe our cavity system.
The simulation framework builds upon the code presented in \cite{GPE_SciPostPhysCodeb}, with relevant extensions for Bragg spectroscopy under an imbalanced transverse pump discussed below. 

\subsection{Cavity Bragg spectroscopy with GPE}

In order to include the on-axis cavity pulse in our simualation, we add a cavity field as drive. This can be taken into account by adding the following part to the Hamiltonian
\begin{equation}
    H_{p} = \hbar \Omega_c (\hat{a} e^{i\omega_{\text{probe}} t} + \hat{a}^{\dagger}e^{-i\omega_{\text{probe}} t}),
\end{equation}
where $\Omega_c$ describes the pumping strength and $\omega_{\text{probe}}$ is the frequency of the probe beam.
After applying the rotating wave approximation and eliminating  the excited state, we derive equations of motion in the Heisenberg picture.
The cavity drive affects only the the equation of motion for the cavity field $a$:

\begin{equation}
    \frac{\partial \hat{a}}{\partial t} = i\Delta_c \hat{a} -i\Omega_c e^{i\delta t} - i \left\{ U_0 B \hat{a} + \frac{1}{2}\sqrt{V_{TP} U_0} \left[\left(\gamma+\frac{1}{\gamma}\right) \Theta_1 + i \left(\gamma-\frac{1}{\gamma}\right)\Theta_2 \right]\right\} -\kappa \hat{a}.
\end{equation}
Here $\Delta_c = \omega_p - \omega_c$, $\Delta_a =\omega_p - \omega_a$, and $\delta = \omega_p - \omega_{\text{probe}}$. $B = \int d^3 r \Psi^{\dagger} \cos^2(k_c x) \Psi $, is the overlap integral of the BEC field operator $\Psi$ with the cavity lattice. $\Theta_1 = \int d^3 r \Psi^{\dagger} \cos(k_p x)\cos(k_c x) \Psi$ and $\Theta_2 = \int d^3 r \Psi^{\dagger} \sin(k_p x)\cos(k_c x) \Psi$ are the integrals of the field operator with the two interference potentials.
In the parameter regime explored in this study ($\kappa \gg E_r$), we can adiabatically eliminate the dynamics of the cavity light field and find the following expression for the cavity light field at the mean-field level ($\hat{a}=\alpha$):

\begin{equation}
    \alpha = \frac{ \frac{1}{2}\sqrt{V_{TP} U_0} \left(  \left(\gamma+\frac{1}{\gamma}\right) \Theta_1 + i \left(\gamma-\frac{1}{\gamma}\right)\Theta_2 \right) + \Omega_c e^{i\delta t}}{\Delta_c - U_0 B + i\kappa}.
\end{equation}

The probing effectively happens through the interference potential, which for the imbalanced pump takes the following form:
\begin{equation}
V = \eta_1 Re(\alpha) \cos(k_p \cdot r) \cos(k_p \cdot r)  + \eta_2 Im(\alpha) \sin(k_c \cdot r) \cos(k_c \cdot r),    \label{eqn_SI:probe_strength}
\end{equation}
where $\eta_{1/2} = \frac{1}{2 } \sqrt{\frac{\VTP g_0^2}{\Delta_a}} (\gamma \pm \frac{1}{\gamma})$.

\subsection{Numerical parameters}
\label{subsection:numpar}
In our numerical simulations based on GPE, we model a $^{87}$Rb BEC cloud with $N=1.4 \times 10^5$ atoms.
The contact interaction strength is $100.4 a_0$, which is mapped to an effective 2D coupling constant, by rescaling it with a harmonic oscillator length given by the trap frequency of the removed dimension. In our case, this length is $a_{\perp} = 0.7312 \times 10^{-6}$m, chosen to match the experimental trap frequency $\omega_z$. 

The simulation parameters were set to closely resemble real experimental conditions. 
The BEC is confined in a harmonic potential with trapping frequencies $\omega_x =2\pi \times \SI{70}{\hertz}$ and $\omega_y = 2\pi \times \SI{89}{\hertz}$. 
The atoms are subjected to potentials created by a transverse pump detuned by $\Delta_a = 2\pi \times 69.8$ GHz from the atomic transition, with an imbalanced factor of $\gamma = 1.45$. 
A cavity is implemented at a $60\degree$ angle to the transverse pump, with a coupling strength $g= 2\pi \times 1.95$ MHz, and dissipation $\kappa = 2\pi \times 147$ kHz. 

The numerical protocol begins by preparing the ground state of the BEC for a given cavity detuning and transverse pump strength. 
The implemented spatial grid extends over $40$ um and consists of $1024$ points. 
The ground state is obtained using $100,000$ imaginary time steps with $dt=-10^{-8} i$. 
Afterward, real-time evolution is performed with a time step of $dt=10^{-7}$. 
The system is then evolved for $3$ ms in the presence of an on-axis cavity probe with a given strength $\Omega_c$ and detuning $\delta$. 
Softening measurements were conducted for three different cavity detunings $\Delta_c = - 2\pi \times (1.5, 3, 6) \SI{}{\mega\hertz}$, and for each of them different probe strengths were applied $\Omega_c = \{17, 35, 30\}\,E_{r}$.
Cavity Bragg spectroscopy was repeated for $50$ different probe detunings $\delta$, sampled in the range $2\pi \times[-5, 5]$ \SI{}{\kilo\hertz}. 
The cavity light during the entire real-time evolution was stored and analyzed following the experimental procedure.

For details of the numerical simulation based on a mean-field model and band structure formalism, see the companion paper\,\cite{suppmat2}. For Fig.\,\ref{fig:concept} (b-e) and Fig.\,\ref{fig5:Syncronization} (c), we used an imbalance parameter $\gamma=1.3$. Note that harmonic confinement, contact interaction and atom-induced dispersive shift is not included in this model.

\subsection{Data analysis}

The collected data were analyzed in a manner analogous to the treatment of experimental data. 
The analysis was based on three key observables: the integrated response in photon number, the response at the probe frequency $R(\delta)$, and the response at the negative probe frequency $R(-\delta)$. 
For the photon number response, the cavity light amplitude was summed and time-averaged for each probe detuning. 
To determine the frequency response, the averaged photon number was plotted against different probe frequencies. 
For the frequency-discriminating observables, a Fast Fourier Transform (FFT) was applied to the cavity light. The absolute values of the frequency spectrum were integrated over the range $\delta\pm 300$ Hz for $R(\delta)$ and $-\delta\pm 300$ for $R(-\delta)$. The results were then plotted as a function of the probe frequency. 
To extract the frequency of the response, either a double-Gaussian or a four-Gaussian peak fitting was applied, depending on whether a single-mode or two-mode response was observed. 

Since the interference potential, which is the effective probing potential depends not only on the cavity probe strength but also on the transverse pump strength, resulting in changes of the system's response when probed at different transverse pump powers. To extract the response (S) we rescale the total response in the analysis by the transverse pump strength, following eq.\,\eqref{eqn_SI:probe_strength}. 

\subsection{Atom Number Dependence and Non-Adiabatic Effects }

To better understand the limitations of the quantitative agreement between the GPE simulations and the experimental data shown in Figs.\,\ref{fig3:Double-roton} and \ref{fig5:Syncronization} of the main text, we consider several contributing factors.

A primary source of uncertainty is the total atom number $N$. In the experiment, $N$ is subject to shot-to-shot fluctuations ($\simeq 15\%$), and more significantly to atom loss during the pump ramp that are not included in the simulations. This loss is expected to be power-dependent: as the final lattice depth increases, atomic densities rise and the confinement tightens, enhancing the system’s sensitivity to both intensity noise and collisional processes. These  effectively reduce the atom number as $V_\mathrm{TP}$ increases and may also lead to a less adiabatic preparation of the final state.  To illustrate the sensitivity of the theoretical prediction to atom number, Fig.~S\ref{fig:GPE_atomnumber} shows the calculated mode energy for two different atom numbers, $N_1 = 1.4 \times 10^5$ and $N_2 = 8 \times 10^4$, using the same parameters as in Fig.~3 of the main text. We find that the critical transverse pump strength $V_\mathrm{TP}^*$ shifts significantly with $N$, and scales approximately as $1/\sqrt{N}$, as expected from the collective nature of the atom-cavity coupling. Even moderate changes in atom number can therefore lead to noticeable shifts in the energy of the modes, limiting quantitative agreement with the experiment. We also note that uncertainties in the calibration of the transverse pump lattice depth are not sufficient to explain the observed discrepancies. 

The most significant discrepancies between theory and experiment are observed beyond the critical point of the SR2 phase. The GPE simulation assumes adiabatic evolution into the ground state, see Sec.\,\ref{subsection:numpar}.  However, crossing the phase boundary, especially in presence of the deep lattice, may excite collective modes in the atomic cloud and trigger residual dynamics that prevent the experimental system from fully relaxing into the ground state. This is particularly relevant near the transition, where the energy gap closes, and the system becomes increasingly sensitive to non-adiabaticity. 
We assign the observed deviations between our numerical simulations and the experimental data to an interplay of these effects — atom number dependent threshold, pump strength dependent loss, and non-adiabaticity — not being captured by the simulation.

\begin{figure}[h]
    \centering
    \includegraphics[width=0.5\linewidth]{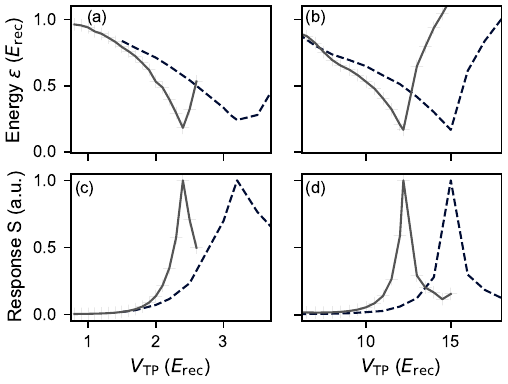}
    \caption{GPE simulation of the mode energy for two different atom numbers: $N_1 = 1.4 \times 10^5$ (solid line) and $N_2 = 8 \times 10^4$ (dashed line). All other parameters are identical to those used in Fig.~3 of the main text.  SR1 [SR2] mode energy $\epsilon$ as a function of pump strength $\VTP$ at cavity detuning $\Delta_c = 2\pi \times -1.5$ \si{\mega\hertz}(a,c) [$\Delta_c = 2\pi \times  -6$ \si{\mega\hertz} (b,d)]. The critical point shifts to higher $V_\mathrm{TP}$ for lower atom number, following approximately a $1/\sqrt{N}$ scaling.}
    \label{fig:GPE_atomnumber}
\end{figure}

\end{appendix}
\end{widetext}
\end{document}